\documentclass[12pt,preprint]{aastex}

\newcommand{\omc}{\mbox{$\omega$ Cen~}}
\def\omc{\hbox{$\omega$ Cen~}}
\def\omcp{\hbox{$\omega$ Cen}}

\shorttitle{On the white dwarf cooling sequence of {\boldmath $\omega$} ~Centauri\altaffilmark{1}} 
\shortauthors{Calamida et al.}

\begin{document}
\title{On the white dwarf cooling sequence of the globular cluster {\boldmath $\omega\,$} Centauri\altaffilmark{1}}

\author{
A. Calamida\altaffilmark{2,3},
C. E. Corsi\altaffilmark{2},
G. Bono\altaffilmark{2,4},
P. B. Stetson\altaffilmark{5},
P. Prada Moroni\altaffilmark{6,7},
S. Degl'Innocenti\altaffilmark{6,7},
I. Ferraro\altaffilmark{2},
G. Iannicola\altaffilmark{2},
D. Koester\altaffilmark{8}, 
L. Pulone\altaffilmark{2}, 
M. Monelli\altaffilmark{9},
P. Amico\altaffilmark{4},
R. Buonanno\altaffilmark{3},
F. Caputo\altaffilmark{2},
S. D'Odorico\altaffilmark{4},
L. M. Freyhammer\altaffilmark{10},
E. Marchetti\altaffilmark{4},
M. Nonino\altaffilmark{11}, and 
M. Romaniello\altaffilmark{4}
}
  \altaffiltext{1}
   {Based on observations collected with the Advanced Camera for Surveys on
    board of the Hubble Space Telescope.}
  \altaffiltext{2}
    {INAF-Osservatorio Astronomico di Roma, via Frascati 33,
    Monte Porzio Catone, Rome, Italy; calamida@mporzio.astro.it
}
  \altaffiltext{3}
    {Universit\`a di Roma Tor Vergata, via della Ricerca Scientifica 1, 00133 Rome,
     Italy; buonanno@roma2.infn.it}
  \altaffiltext{4}{European Southern Observatory, Karl-Schwarzschild-Str. 2,
     D-85748 Garching bei Munchen, Germany; pamico@eso.org, sdodoric@eso.org,
     emarchet@eso.org, mromanie@eso.org}
  \altaffiltext{5}{Dominion Astrophysical Obs., HIA, NRC, 5071 West Saanich Road, 
     Victoria BC V9E~2E7, Canada; peter.stetson@nrc-cnrc.gc.ca}
  \altaffiltext{6}
  {Dipartimento di Fisica "E. Fermi", Univ. Pisa, Largo B. Pontecorvo 2, 56127 Pisa
    Italy; prada@df.unipi.it, scilla@df.unipi.it}
  \altaffiltext{7}{INFN, Sez. Pisa, via Largo B. Pontecorvo 2, 56127 Pisa, Italy}
  \altaffiltext{8}{Institut für Theoretische Physik und Astrophysik, University of Kiel, 24098 Kiel, Germany; koester@astrophysik.uni-kiel.de}
  \altaffiltext{9}{IAC - Instituto de Astrofisica de Canarias, Calle Via Lactea,
   E38200 La Laguna, Tenerife, Spain; monelli@iac.es}
  \altaffiltext{10}{Centre for Astrophysics, University of Central Lancashire,
   Preston PR1 2HE; lmfreyhammer@uclan.ac.uk}
   \altaffiltext{11}{INAF-Osservatorio Astronomico di Trieste, via G.B. Tiepolo 11,
    40131 Trieste, Italy; nonino@oats.inaf.it}

\date{\centering drafted \today\ / Received / Accepted }

\begin{abstract}
We present deep and precise photometry ($F435W$, $F625W$, $F658N$) 
of \omc collected with the Advanced Camera for Surveys (ACS) on board the 
Hubble Space Telescope (HST). We have identified $\approx$ 6,500 white dwarf 
(WD) candidates, and the ratio of WD and Main Sequence (MS) star counts is 
found to be at least a factor of two larger than the ratio of CO-core WD 
cooling and MS lifetimes. This discrepancy is not explained by the possible 
occurrence of a He-enhanced 
stellar population, since the MS lifetime changes by only 15\% when 
changing from a canonical (Y=0.25) to a He-enhanced composition (Y=0.42). 
The presence of some He-core WDs seems able to explain the observed star 
counts. The fraction of He WDs required ranges from 10\% to 80\% depending 
on their mean mass and it is at least five times larger than 
for field WDs. 
The comparison in the Color Magnitude Diagram between theory and observations 
also supports the presence of He WDs. Empirical evidence indicates that 
He WDs have been detected in stellar systems hosting a large sample of 
extreme horizontal branch stars, thus suggesting that a fraction of red 
giants might avoid the He-core flash.  
\end{abstract}



\section{Introduction}

White Dwarfs (WDs) in Galactic Globular clusters (GGCs) represent the intersection 
of several theoretical and empirical astrophysical problems (Koester 2002;
Hansen \& Liebert 2003; Hansen 2004). They play a crucial role in constraining 
the correctness of the physical assumptions adopted to construct 
WD evolutionary models (Prada Moroni \& Straniero 2007). 
Cluster WDs possess several advantageous features. 
{\em Homogeneous sample --} Cluster WDs are located at the same 
distance and generally have about the same reddening. Moreover, at all luminosities
the colors of cluster WDs are systematically
bluer than Main Sequence (MS) stars. This means that to properly
identify cluster WDs we can use a Color Magnitude Diagram (CMD) instead of a color-color diagram. Therefore, 
cluster WDs are not affected by the color
degeneracy with MS stars that affects field WDs (Hansen \& Liebert 2003).
{\em Statistics --} Evolutionary predictions indicate that in a
GC with an age of 12 Gyr and a Salpeter-like initial mass function 
the number of WDs should be three orders of magnitude 
larger than the number of Horizontal Branch (HB) stars (Brocato et al.\ 1999). 
This together with the high stellar concentration implies that 
the expected density of WDs in GCs is several 
orders of magnitude larger than in the Galactic field. 
{\em Origin --} For cluster WDs we can trace back the evolutionary
properties of the progenitors, since both the cluster age and the 
chemical composition are well known (Kalirai et al.\ 2007, KA07). 

However, cluster WDs also present a few drawbacks. 
{\em Crowding --} they are faint objects severely affected by crowding 
problems (Moehler et al.\ 2004).
{\em Cluster vs Field --}  Current data do not allow us to establish 
definitively whether cluster WDs are the analogs of field WDs. The impact that 
the high density environment of GCs may have on the formation and evolution of 
cluster WDs is still poorly known (Monelli et al.\ 2005, MO05).

In a previous investigation based on three out of nine ACS 
pointings we have already addressed the properties of WDs in \omc (MO05).  
In the meantime, deep photometric investigations called attention to the 
occurrence of a split 
along the MS of \omc (Bedin et al.\ 2004). Spectroscopic data have
indicated that the stars distributed along the bluer MS (30\% of 
the entire population) are $\approx$ 0.3 dex more metal-rich than 
the typical population and probably also He-enriched (Y$\sim$ 0.42, 
Piotto et al.\ 2005).  

\section{Observations and data reduction}

We use archival multiband ($F435W$, $F625W$, $F658N$) photometric data collected
with the ACS on board the HST. The current data set includes eight out of the 
nine pointings located across the cluster center that have already been discussed 
by Castellani et al.\ (2007, see their Fig.~1, CA07). The central pointing of 
the $3\times 3$ mosaic was omitted due to the severe 
crowding of the innermost regions. For each field, the $F435W$- and $F625W$-band 
data consist of one shallow ($8$s) and three deep ($340$s each) 
exposures, while the $F658N$-band data consist of four exposures
of $440$s each. The raw frames were pre-reduced by the 
standard HST pipeline. The entire set of images was reduced simultaneously 
with {\tt DAOPHOT~IV/ALLFRAME} and the final catalog includes 
more than one million stars. The photometry was kept in the Vega 
system (Sirianni et al.\ 2005).

Fig.~1 (left) shows the $F435W,$ $F435W-F625W$ CMD for selected 
cluster stars. From this catalog we selected all the stars 
systematically bluer than MS stars and fainter than extreme 
HB (EHB) stars ($B \lesssim 20$, see solid black lines in the 
left panel of Fig.~1). We ended up with a sample of $\approx 60,000$ 
stars.  The photometry of these stars was performed 
once again using {\tt ROMAFOT}, but only for the deep exposures. 
Individual stars have been interactively checked in every image, 
and the WD candidates were measured either as isolated stars 
or together with neighbor stars. A significant fraction 
of the originally selected detections turned out to be either cosmic 
rays or spurious detections close to saturated stars, or detections 
too faint to be reliably measured on individual images. Fig.~1 also 
shows the $F435W,$ $F435W-F625W$ (middle) and the $F435W,$ $F435W-F658N$ 
(right) CMDs based, this time, on the {\tt ROMAFOT}  photometry. 
Data plotted in these panels show that the candidate cluster 
WDs ($\sim 6,500$) are distributed along a well defined 
star sequence fainter than EHB stars and bluer 
than MS stars (MO05). To our knowledge this is the largest sample 
of cluster WD candidates ever detected. The current sample is 
$\approx$ 40\% of the WDs identified in all GGCs combined
(Hansen et al.\ 2004,2007) and $\approx$ 50\% of all spectroscopically 
confirmed field WDs (Eisenstein et al.\ 2006).   
 
\section{Results and final remarks}

In order to compare theory and observation we have adopted the cooling 
sequences for CO-core and H-rich envelopes (DA WDs) by  Althaus \& 
Benvenuto (1998), for CO-core and He-rich envelopes (DB WDs)
by Benvenuto \& Althaus,(1997), and for He-core (He WDs) 
by Serenelli et al.\ (2002). The theoretical predictions were 
transformed into the observational plane by adopting the 
pure H and pure He WD atmosphere models computed by 
Koester et al.\ (2005). Predicted magnitudes in the Vega system were computed 
using the ACS bandpasses ({\tt ftp://ftp.stsci.edu/cdbs/cdbs1/comp/acs}), 
while their zero-points are based on the Vega spectrum 
(Bohlin \& Gilliland 2004, {\tt ftp://ftp.stsci.edu/cdbs/cdbs2/calspec/}).  
Fig.~2 shows the comparison, in the $F435W, F435W-F625W$ CMD, 
between the candidate cluster WDs and predicted cooling sequences for 
DA (top, CO-core $+$ H envelope), DB (middle, CO-core $+$ He envelope), 
and He (bottom, He-core) WDs. Note that in this figure we have plotted only
stars with $\sigma(F435W-F625W)\le 0.3$, i.e., objects above a 5-$\sigma$ 
detection threshold. Predicted cooling sequences were plotted for 
a true \omc distance modulus $\mu_0 = 13.70\pm0.06$ 
(Del Principe et al.\ 2006) and a reddening 
$E(B-V)=0.11\pm 0.02$ (Calamida et al.\ 2005). Using the reddening law 
from Cardelli et al.\ (1989) and $R_V=3.1$, we find $A_{F435W}=0.46$, 
$E(F435W-F625W)=0.17$, and $E(F435W-F658N)=0.18$.    
Data plotted in this figure further strengthen the preliminary evidence 
brought forward by MO05 based on a smaller WD sample:  DA and DB WD 
cooling sequences are, at fixed magnitude, systematically bluer (hotter) 
than the bulk of candidate cluster WDs. On the other hand, the predicted 
He WD cooling sequences are consistent with a substantial fraction of the 
observed candidate WDs. 
The discrepancy between predicted DA/DB WDs and observations can hardly 
be caused by the adopted WD atmosphere models. The difference in 
color at fixed magnitude between the same cooling sequences transformed 
using WD models provided by Bergeron, Wesemael, \& Beauchamp (1995) 
and our WD models is, for $T_{eff} \ge 10,000$, at most 
$\sim 0.02\,$mag\footnote{The $F625W$- and $F658N$-band 
WD cooling sequences plotted in Figs. 3,4 of MO5 were unknowingly 
interchanged. This mismatch does not affect the conclusions of 
that investigation.}.  

To further constrain this circumstantial empirical evidence, 
MO05 compared the ratio between observed WD and HB star counts with the 
corresponding evolutionary times. However, CA07 found an apparent excess
of HB stars in \omc based on a photometric data set covering almost 
the entire cluster.  In order to avoid deceptive uncertainties in the 
observed ratios we therefore decided to use MS stars located across 
the turnoff (TO) region as our reference sample, rather than HB stars.
Specifically, we employ the magnitude bin $18.775 \le B \le 19.025$ mag 
(see the green box in Fig.~1) because theory and observations suggest that 
its population depends minimally on the initial mass function (Zoccali \& 
Piotto 2000) since these stars represent a very small range of mass (CA07).
Moreover, this magnitude range is minimally affected by completeness problems. 
The MS star counts are 
based on the {\tt ALLFRAME} catalog, while the WD star counts 
are based on the {\tt ROMAFOT} catalog. The former data set is complete
along the MS, while the latter is less contaminated by spurious detections 
in the WD region. The WDs were selected in three different magnitude bins: 
$B \le 24$, 24.5, and 25 (purple, cyan, and red dots in the middle and right
panels of Fig.~1) to trace the sample completeness when considering 
fainter magnitudes. We did not apply any selection criteria to estimate the 
star counts, apart from the magnitude limits.  
We found that the observed ratios in the two different CMDs (see 
lines 1, 2 in Table~1) agree quite well in the brighter magnitude bin
($0.052\pm0.002$ vs $0.050\pm0.002$) while they steady decrease in the 
diagram based upon the narrow $F658N$ bandpass when moving 
toward fainter magnitudes ($0.163\pm0.004$ vs $0.147\pm0.004$). This effect 
is due to the difference in the completeness between the wide $F625W$ 
and the narrow $F658N$ band, the latter obviously being shallower.  
In order to constrain the dependence on the adopted MS sample, we 
counted MS stars once again in the box $19.025 \le B \le 19.275$ mag. 
Ratios listed in Table 1 show that the difference in the brighter bin 
is on average $\approx 16$\%.  

In order to define the typical stellar mass of MS turn-off stars 
we adopted two cluster isochrones for $t=12$ Gyr (see top panel in Fig.~2) 
with canonical primordial helium content Y=0.25 (Spergel et al.\ 2007) 
and metal abundances (Z=0.0004, Z=0.0015) that bracket the observed 
spread in metallicity of \omc stars (Sollima et al.\ 2005; 
Villanova et al.\ 2007). These isochrones were transformed into the 
observational plane using the atmosphere models provided by 
Brott \& Hauschildt (2005).  
The above isochrones are based on evolutionary tracks computed 
with an updated version of the FRANEC evolutionary code 
(Chieffi \& Straniero 1989) including microscopic element
diffusion (Cariulo et al.\ 2004; CA07). 
In order to estimate the lifetime spent by MS stars in crossing 
the specified magnitude bin we constructed two evolutionary tracks with 
$M/M_\odot$=0.75 (Z=0.0004) and $M/M_\odot$=0.78 (Z=0.0015). We found that 
the average crossing time for these two tracks is $\approx$ 950 Myr. 
The predicted lifetime ratios between DA/DB WD ($M_{WD}$=0.5-0.9 $M_\odot$) and 
MS ($M_{MS}$=0.75-0.78 $M_\odot$) stars attain, within the uncertainties, 
quite similar values (see lines 3 to 6 in Table~1). The errors in the 
lifetime ratios include an uncertainty of $\approx$ 10\% in the adopted 
input physics (CA07). The ratios for He WDs are, as expected, larger and 
in the brighter bin ($F435W\le 24$) they are at least a factor of three 
larger than CO-core ratios (see lines 7 to 9 in Table~1).  
To estimate possible uncertainties in distance and in the reddening 
correction, we adopted a larger apparent distance modulus 
($DM_B$=14.36 vs 14.16). The difference in the two sets of lifetime 
ratios for $M_{WD}$=0.5 is at most 18\%.  

This comparison between theory and observation indicates that 
the star count ratios in the brighter magnitude bin (see Fig.~3) are at 
least a factor of two larger than predicted by DA/DB WD cooling 
times. On the other hand, the observed ratios are at least a factor
of four smaller than predicted by He WD cooling times. 
The discrepancy between the observed WD star counts and the predicted CO-core 
ratios can hardly be explained by incompleteness problems affecting 
the sample of candidate WDs, which would go in the direction 
of increasing the discrepancy between theory and observation.
The same conclusion would apply for a putative increase in the 
mean mass of CO-core WDs, since the lifetime ratios on average 
decrease---as expected---by at least a factor of two 
(see lines 4, 6 in Table~1). An increase in the mean mass of He WDs, 
on the other hand, does not imply a steady decrease in the cooling 
lifetime (see lines 7 to 9 in Table~1). This nonlinear behavior is 
due to the occurrence of CNO thermonuclear flashes in the mass range 
$0.22 \lesssim M/M_\odot \lesssim 0.422$ (Serenelli et al.\ 2002).         
The lifetime ratios quoted above indicate that predicted He WD lifetimes 
are at least a factor of two larger than observed.  

The quoted discrepancies are also minimally affected by a decrease 
in the cluster age of 2 Gyr (see middle panel of Fig.~2). We constructed 
two evolutionary tracks with $M/M_\odot=0.77$ (Z=0.0004) and 
$M/M_\odot=0.80$ (Z=0.0015) and found that the mean time they 
spend in crossing the specified magnitude bin is only 10\% shorter than 
for the older models (see lines 10 to 14 in Table~1). Therefore, this 
decrease in the assumed age hardly affects the discrepancy between theory and 
observations.  As another attempt to account for our findings we also 
considered a possible increase in the He content of MS stars. In particular, 
we adopted two cluster isochrones with same age and metal abundances
as the canonical ones, but with a He-enhanced ($Y=0.42$) composition 
(see bottom panel of Fig.~2). In order to represent a possible 
spread in He content, we estimated the predicted ratios for
a mix of stellar populations consisting of 70\% stars with canonical 
He content (Y=0.25) and 30\% He-enhanced stars. 
We constructed two He-enhanced evolutionary tracks with 
$M/M_\odot$=0.55 (Z=0.0004) and $M/M_\odot$=0.57 (Z=0.0015) 
and found that the mean lifetime they spend in the specified 
magnitude bin is 460 Myr. Therefore, the MS lifetimes of the 
mixed-He population decrease by only 
$\sim 15$\% (800 vs 950 Myr) relative to the canonical population. 
Data plotted in Fig.~3 (see also Table~1) indicate that the 
occurrence of a He-enhanced sub-population in \omc can not by itself explain 
the discrepancy between the star counts and the lifetime ratios. 

Let us assume, as a working hypothesis, that the current sample of 
candidate cluster WDs represents a mix of CO-core and He-core WDs. 
The aforementioned lifetimes suggest that the fraction of He WDs 
ranges from 80\% (if we assume a mean mass of $0.5 M_\odot$ for 
the CO-core and $0.3 M_\odot$ for the He-core WDs) to 10\% (for 
a mean mass of $0.5 M_\odot$ for the CO-core and $0.23 M_\odot$ 
for the He-core WDs). The latter fraction decreases further if we 
assume still smaller He-core WDs, but current empirical estimates indicate 
that the lower limit ranges from $\approx 0.17$ to $\approx 0.2\; M_\odot$ 
(Moehler et al.\ 2004; Kepler et al.\ 2007). This evidence, if supported 
by independent spectroscopic measurements, indicates that cluster 
WD samples might present different intrinsic properties when compared 
with field WDs. Current estimates based on the large SDSS sample of WDs indicate 
that only 2\% of field DA WDs possess masses smaller than $0.45 M_\odot$.    
Note that the current fraction of He-core WDs is different by only a factor of 2-3 from the
global binary frequency in \omc (Mayor 1996) and in good agreement with the binary 
fraction ($\approx$ 10\%) in GCs in general (Davies et al.\ 2006).
A similar excess of He WDs in the old open cluster NGC~6791 was proposed by
KA07.  They found that roughly 40\% of the WDs in this system did not experience 
the expected core-helium flash at the tip of the red giant branch (RGB). 
These objects end their evolution as He-core WDs after having lost a significant 
fraction of their envelope. According to evolutionary prescriptions they are 
the aftermath of an extreme mass loss episode  possibly caused either by stellar 
collisions or by close binary interactions (Castellani et al.\ 2006a,b, CA06a, CA06b).  
However, Bedin et al.\ (2005), using deep ACS images, suggested 
that the color distribution of WDs in NGC~6791 does not support the 
occurrence of He WDs. 

The available observations present, as suggested by the referee, puzzling
empirical aspects. Detailed photometric investigation of WDs in GCs like M4
(Hansen et al.\  2004) and NGC~6397 (Hansen et al.\ 2007) do not show evidence
for He-core WDs. On the other hand, Sandquist \& Martel (2007) found a well
defined deficiency ($\approx$ 20\%) of bright RGs in NGC~2808
and suggested that the missing giants might produce He-core WDs. An enhanced
mass loss efficiency, driven by metal content, was suggested by KA07 
to account for He-core WDs in NGC~6791. However, the peak in the
metallicity distribution of \omc (Kayser et al.\ 2006) and the metallicity
of NGC~2808 (Carretta 2006) are at least 1.5 dex more metal-poor than 
NGC~6791. A possible He enrichment has been proposed to account for EHB stars 
and the complex MS structure in \omc and in NGC~2808 (D'Antona et al.\ 2002;
Piotto et al.\ 2007). If we assume a canonical helium-to-metal enrichment 
ratio ($\Delta Y/\Delta Z\approx2.5$) a similar enhancement could also be 
present in NGC~6791. However, WD/MS and HB/MS (CA07) 
star count ratios in \omc do not seem to support this hypothesis: for a canonical 
enrichment ratio, the putative He-rich stars in \omc should have above Solar 
metallicities. A single simple hypothesis of stellar evolution
driven by cluster structural parameters and dynamical evolution can hardly 
account for the quoted He WD identifications, since the central density 
in \omc and in NGC~2808 is significantly larger than in NGC~6791. 
Therefore, we are left with compelling evidence
that He WDs have been detected/predicted in stellar systems that host
sizable samples of EHB stars (\omcp, CA07; NGC~2808, CA06a; NGC6791,
KA07). However, the natural progeny of EHB stars are
CO-core WDs. The He enrichment scenario can account for 
EHB stars, but does not explain, for canonical mass loss rates, the 
occurrence of He WDs. On the other hand, if a substantial fraction of RGs avoids 
the He-core flash, they will end up their evolution, according to the 
residual envelope mass, either as EHB/CO-core WDs or as He-core WDs 
(Hansen 2005; CA06b).



\clearpage
\tablewidth{0pt}
\begin{deluxetable}{lllllll}
\tabletypesize{\scriptsize}

\tablecaption{
Ratios between CO and He WD cooling times and MS lifetimes}\label{tbl-1}
\tablehead{
\multicolumn{1}{c}{$F435W$} & \multicolumn{2}{c}{24.0} & \multicolumn{2}{c}{24.5} & \multicolumn{2}{c}{25.0}  
}
\startdata
$N_{WD}/N_{MS}$ & 0.052(2)\tablenotemark{a}&0.044(2)\tablenotemark{b} & 0.095(2)\tablenotemark{a}&0.080(2)\tablenotemark{b} & 0.163(4)\tablenotemark{a}&0.137(3)\tablenotemark{b} \\
$N_{WD}/N_{MS}$ & 0.050(2)\tablenotemark{c}&0.042(2)\tablenotemark{d} & 0.090(3)\tablenotemark{c}&0.076(3)\tablenotemark{d} & 0.147(4)\tablenotemark{c}&0.124(3)\tablenotemark{d} \\
DA[0.5]\tablenotemark{e} & 0.021(3)\tablenotemark{f}&0.019(2)\tablenotemark{g}   & 0.048(7)\tablenotemark{f}&0.044(6)\tablenotemark{g} & 0.12(2)\tablenotemark{f}&0.11(2)\tablenotemark{g}   \\
DA[0.9]\tablenotemark{e} & 0.0045(6)\tablenotemark{f}&0.0029(4)\tablenotemark{g} & 0.029(4)\tablenotemark{f}&0.018(2)\tablenotemark{g} & 0.08(1)\tablenotemark{f}&0.08(1)\tablenotemark{g}   \\
DB[0.5]\tablenotemark{e} & 0.021(3)\tablenotemark{f}&0.020(2)\tablenotemark{g}   & 0.057(8)\tablenotemark{f}&0.047(7)\tablenotemark{g} & 0.13(2)\tablenotemark{f}&0.13(2)\tablenotemark{g}   \\
DB[0.9]\tablenotemark{e} & 0.0040(5)\tablenotemark{f}&0.0034(5)\tablenotemark{g} & 0.016(2)\tablenotemark{f}&0.008(1)\tablenotemark{g} & 0.07(1)\tablenotemark{f}&0.050(7)\tablenotemark{g}  \\
He[0.23]\tablenotemark{e} & 0.35(5)\tablenotemark{f}&0.38(5)\tablenotemark{g}     & 0.51(7)\tablenotemark{f}&0.57(8)\tablenotemark{g}  & 0.70(10)\tablenotemark{f}&0.80(11)\tablenotemark{g} \\
He[0.3]\tablenotemark{e}  & 0.07(1)\tablenotemark{f}&0.040(6)\tablenotemark{g}    & 0.18(3)\tablenotemark{f}&0.17(2)\tablenotemark{g}  & 0.33(5)\tablenotemark{f}&0.34(5)\tablenotemark{g}   \\
He[0.45]\tablenotemark{e} & 0.15(2)\tablenotemark{f}&0.15(2)\tablenotemark{g}     & 0.32(4)\tablenotemark{f}&0.30(4)\tablenotemark{g}  & 0.63(9)\tablenotemark{f}&0.61(9)\tablenotemark{g}   \\

DA[0.5]\tablenotemark{e} & 0.019(3)\tablenotemark{h}&0.024(3)\tablenotemark{i}   & 0.044(6)\tablenotemark{h}&0.057(8)\tablenotemark{i} & 0.11(2)\tablenotemark{h}&0.14(2)\tablenotemark{i}  \\
DB[0.5]\tablenotemark{e} & 0.019(3)\tablenotemark{h}&0.024(3)\tablenotemark{i}   & 0.052(7)\tablenotemark{h}&0.067(9)\tablenotemark{i} & 0.12(2)\tablenotemark{h}&0.16(2)\tablenotemark{i}  \\
He[0.23]\tablenotemark{e}& 0.32(5)\tablenotemark{h}&0.42(6)\tablenotemark{i}     &  0.47(7)\tablenotemark{h}&0.61(8)\tablenotemark{i}  & 0.64(9)\tablenotemark{h}&0.82(2)\tablenotemark{i}  \\
He[0.3]\tablenotemark{e} & 0.07(1)\tablenotemark{h}&0.08(1)\tablenotemark{i}     &  0.17(2)\tablenotemark{h}&0.21(3)\tablenotemark{i}  & 0.30(4)\tablenotemark{h}&0.39(5)\tablenotemark{i}  \\
He[0.45]\tablenotemark{e}& 0.14(2)\tablenotemark{h}&0.18(3)\tablenotemark{i}     &  0.29(4)\tablenotemark{h}&0.37(5)\tablenotemark{i}  & 0.58(8)\tablenotemark{h}&0.75(11)\tablenotemark{i} \\
\enddata                                                                                     
\tablenotetext{a}{Star counts based on the $F435W, F435W-F625W$ CMD and the MS box at $B=18.90$. 
Numbers in parentheses are the uncertainties on the last decimal figures(s).} 
\tablenotetext{b}{Star counts based on the $F435W, F435W-F625W$ CMD and the MS box at $B=19.15$.}  
\tablenotetext{c}{Star counts based on the $F435W, F435W-F658N$ CMD and the MS box at $B=18.90$.}
\tablenotetext{d}{Star counts based on the $F435W, F435W-F658N$ CMD and the MS box at $B=19.15$.}
\tablenotetext{e}{Predicted cooling and lifetime ratios for CO-core (DA,DB) and He-core WDs. Numbers in square brackets 
are the WD masses in solar units.} 
\tablenotetext{f}{Estimates based on a distance modulus of $DM_B=14.16$.}   
\tablenotetext{g}{Estimates based on a distance modulus of $DM_B=14.36$.}   
\tablenotetext{h}{Estimates based on a cluster age of 10 Gyr.} 
\tablenotetext{i}{Estimates based on 70\% He-normal (Y=0.25) and 30\% He-enriched (Y=0.42) stars.}
\end{deluxetable}

\clearpage

\begin{figure}
\includegraphics[height=0.75\textheight,width=1.00\textwidth]{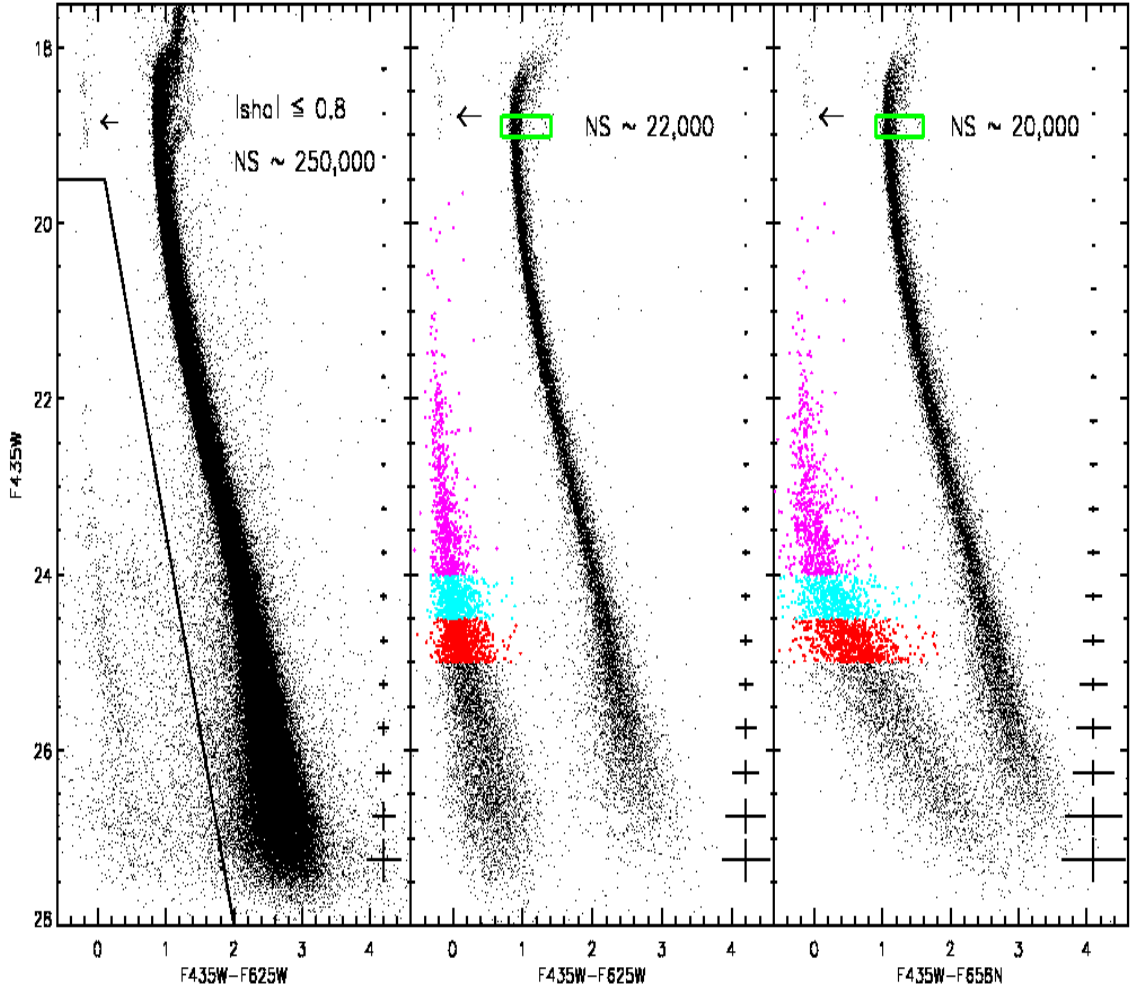}
\vspace{-2.0truecm}
\caption[]{Left -- F435W, F435W-F625W CMD based on data collected with ACS@HST and 
reduced with {\tt ALLFRAME}. Stars plotted in this diagram were 
selected according to {\em sharpness} and {\em separation}. The solid black lines 
show magnitude and color ranges of the candidate cluster WDs. The arrow marks 
EHB stars, while the error bars on the right display intrinsic errors in magnitude 
and color. Middle -- F435W, F435W-F625W CMD based on deep images collected with 
ACS@HST and reduced with {\tt ROMAFOT}. The MS and the TO regions are less populated 
because only the stars located close to candidate WDs have been measured. 
Moreover, stars with $F435W\lesssim 18$ are saturated in deep images. The green 
box shows TO region adopted for MS star counts. Candidate WDs with $F435W\le 24$, 
24.5, and 25 are marked with different colors. Right -- Same as the middle, but for 
the F435W, F435W-F658N CMD.} 
\label{fig:apjfig1} 
\end{figure}

\begin{figure}
\includegraphics[height=0.80\textheight,width=0.65\textwidth]{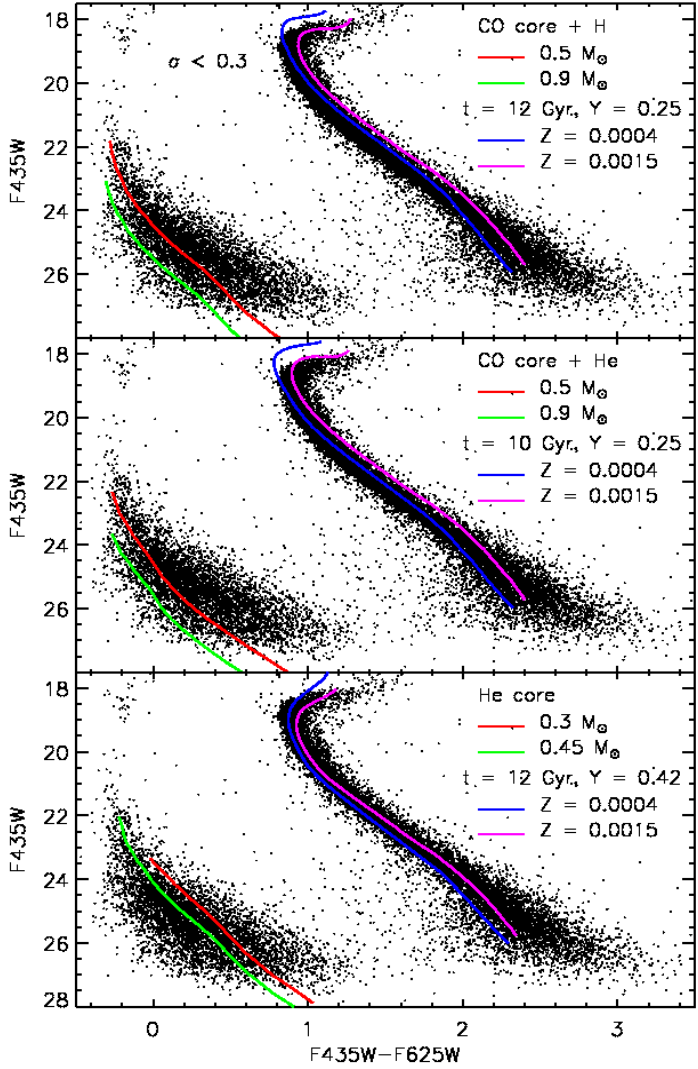}
\vspace{-2.0truecm}
\caption[]{Top -- F435W, F435W-F625W CMD, only stars with 
$\sigma(F435W-F625W)\le 0.3$ are plotted. The green and the red line 
show the cooling sequences for selected DA (CO-core + H envelope) WDs, while 
the blue and the purple line are two cluster isochrones for $t=12$ Gyr and
different metal abundances (see labeled values). Middle -- same as the top, 
but the cooling sequences refer to DB (CO-core + He envelope) WDs and two
isochrones for $t=10$ Gyr. Bottom -- same as the top, but the cooling 
sequences refer to He-core WDs and two isochrones for $t=12$ Gyr and a 
He-enhanced (Y=0.42) chemical composition.} 
\label{fig:apjfig2}
\end{figure}


\begin{figure}
\begin{center}
\includegraphics[height=0.45\textheight,width=0.6\textwidth]{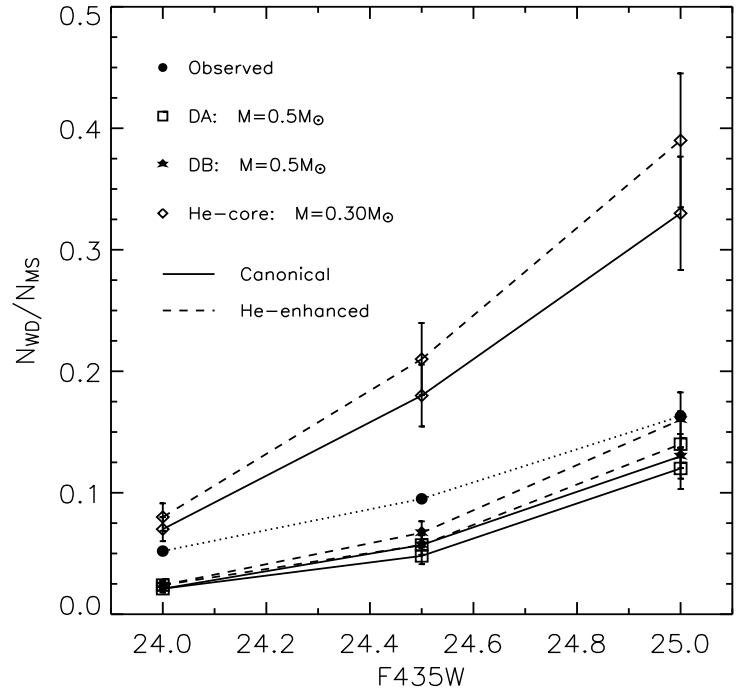}
\caption[]{Star count ratios ($N_{WD}/N_{MS}$) and predicted ratios between 
WD cooling times and MS lifetimes versus $F435W$ magnitude. Solid and dashed lines 
show the ratios for He-normal and He-enhanced structures.} 
\label{fig:apjfig3}
\end{center}
\end{figure}

\end{document}